\documentclass[11pt,twoside]{article} 
\usepackage{asp2004}
\usepackage{epsf}
\usepackage{amsmath,amssymb}
\usepackage{lscape} 

\begin{document} 
\title{Mass-loss and multicomponent flow from central stars of planetary
nebulae}
\author{J. Krti\v{c}ka,}
\affil{Masaryk University, Kotl\'a\v rsk\'a 2,
 CZ-611 37 Brno, Czech Republic}
\author{J. Kub\'at}
\affil{Astronomick\'y \'ustav AV \v{C}R, CZ-251 65 Ond\v{r}ejov, Czech
Republic}
\begin{abstract} 
We calculate multicomponent radiatively driven stellar wind models
suitable for central stars of planetary nebulae.
Some of these stellar winds may be adequately modelled using
one-component models, however for some of them multicomponent models
are necessary.
We obtain a range of stellar parameters for different types of
mass-loss.
\end{abstract}

\section{Introduction}
Hot star winds are accelerated mainly by absorption of radiation in
resonance lines. This process can be divided into
two steps: 
\begin{enumerate}
\itemsep -1pt
\item
transfer of light momentum to C, N, O, Fe, etc., by absorption of
radiation,
and by Thomson
scattering
to
free electrons,
\item transfer of obtained momentum to predominant wind component (H,
He).
\end{enumerate}
Since the acceleration of different wind species may be different, their
wind velocities may differ and so hot star winds have a multicomponent
nature.
For {\em high density} winds (e.g. of galactic O stars) this
multicomponent nature does not influence
the
wind structure,
however for {\em low density} winds multicomponent effects occur, for
example frictional heating, decoupling of wind components, etc.
(see Krti\v{c}ka \& Kub\'at 2001).
To estimate the importance of multicomponent effects for the winds of
central stars of planetary nebulae we calculate multicomponent wind
models for these stars.

\section{Model assumptions}

Basic assumptions of our models are the following:
\begin{itemize}
\itemsep -1pt
\item
we assume a stationary spherically-symmetric flow,
\item we solve the continuity, momentum and energy equations for each
component of the flow, namely for absorbing ions, nonabsorbing ions
(hydrogen and helium), and electrons (see Krti\v{c}ka \& Kub\'at 2001),
\item we assume solar chemical composition,
\item line radiative force is calculated in the CAK approximation
(Castor, Abbott \& Klein 1975),
we neglect wind instabilities 
and magnetic fields. 
\end{itemize}

\section{Multicomponent model equations}

For each component $a$ of the flow (i.e. accelerated ions, passive
component (hydrogen and helium) and electrons) we solve continuity,
momentum and energy equations in the form of
\begin{gather}
\frac{1}{r^2}\frac{\text{d}}{\text{d} r}\left(r^2\rho_a{v_r}_a\right)=0, \\
{v_r}_a\frac{\text{d} {v_r}_a}{\text{d} r}=
{g}_{a}^{\mathrm{rad}}-g-\frac{1}{{\rho}_a}\frac{\text{d}}{\text{d} r}\left({a}_a^2{\rho}_a\right)
 +\frac{q_a}{m_a}E +
\sum_{b\neq a} 
{g}_{ab}^{\mathrm{fric}}, \\
\frac{3}{2}k{v_r}_a\frac{\rho_a}{m_a}\frac{\text{d} T_a}{\text{d} r}+
\frac{a_a^2\rho_a}{r^2}\frac{\text{d}}{\text{d} r}\left(r^2 {v_r}_a\right)=
Q_a^{\mathrm{rad}}+
\sum_{b\neq a}\left(
Q_{ab}^{\mathrm{ex}}+
Q_{ab}^{\mathrm{fric}}\right), 
\end{gather}
where $\rho_a$ is the density of a component $a$, ${v_r}_a$ is
radial velocity, ${a}_a$ is the isothermal sound speed,
${g}_{a}^{\mathrm{rad}}$ is the radiative acceleration either due to the
line-transitions or due to the electron scattering, $E$ is the electric
polarisation field, ${g}_{ab}^{\mathrm{fric}}$ is the frictional
acceleration, $T_a$ is the temperature,
$Q_a^{\mathrm{rad}}$ is the radiative heating/cooling
term
(calculated using electron thermal balance method, Kub\'at et al. 1999),
$Q_{ab}^{\mathrm{ex}}$ is the heat exchange and
$Q_{ab}^{\mathrm{fric}}$ is the frictional heating.

\section{The frictional acceleration}

The frictional acceleration ${g}_{ab}^{\mathrm{fric}}$ depends on the
velocity difference via the Chandrasekhar function
(${g}_{ab}^{\mathrm{fric}}\sim G(x_{ab})$),

\begin{equation}
G(x_{ab})=\frac{{\Phi(x_{ab})-x_{ab}\frac{\displaystyle\text{d}\Phi(x_{ab})}{\displaystyle\text{d}
x_{ab}}}}{2
x_{ab}^2},
\end{equation}
where $\Phi(x_{ab})$ is the error-function and the relative velocity
difference between wind components $a$ and $b$ is
\begin{equation}
x_{ab}=\frac{|{v_r}_b-{v_r}_a|}{\alpha_{ab}}.
\end{equation}
%

\begin{figure}
\begin{center}
\resizebox{0.49\textwidth}{!}{\includegraphics{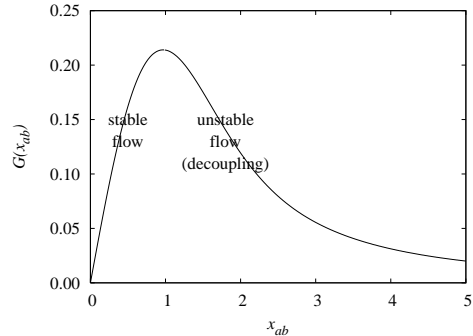}}
\end{center}
\caption{The run of Chandrasekhar function $G(x_{ab})$. The
frictional acceleration is proportional to the Chandrasekhar function,
${g}_{ab}^{\mathrm{fric}}\sim G(x_{ab})$.}
\label{chandra}
\end{figure}

For small relative velocity differences, $x_{ab}\lesssim1$, the
Chandrasekhar function $G(x_{ab})$ is increasing, the flow is stable in
this case, however for larger velocity differences, $x_{ab}\gtrsim1$,
$G(x_{ab})$ is decreasing (see Fig.\ref{chandra}).
The latter
behaviour enables {\em decoupling} of wind components, the flow is
unstable for larger velocity differences (Owocki \& Puls 2002,
Krti\v{c}ka \& Kub\' at 2002).

\section{Examples of calculated wind models}



\begin{figure}
\begin{center}
\resizebox{0.49\hsize}{!}{\includegraphics{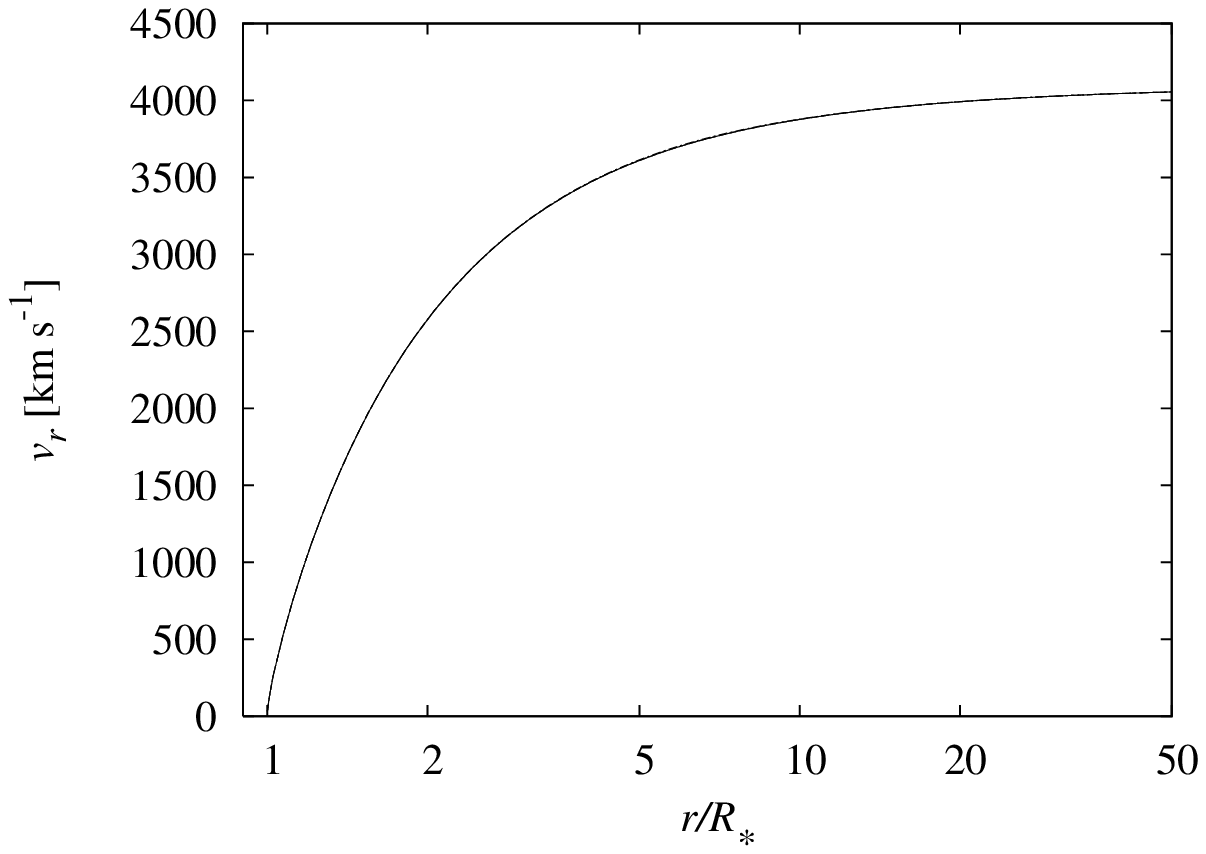}}
\resizebox{0.49\hsize}{!}{\includegraphics{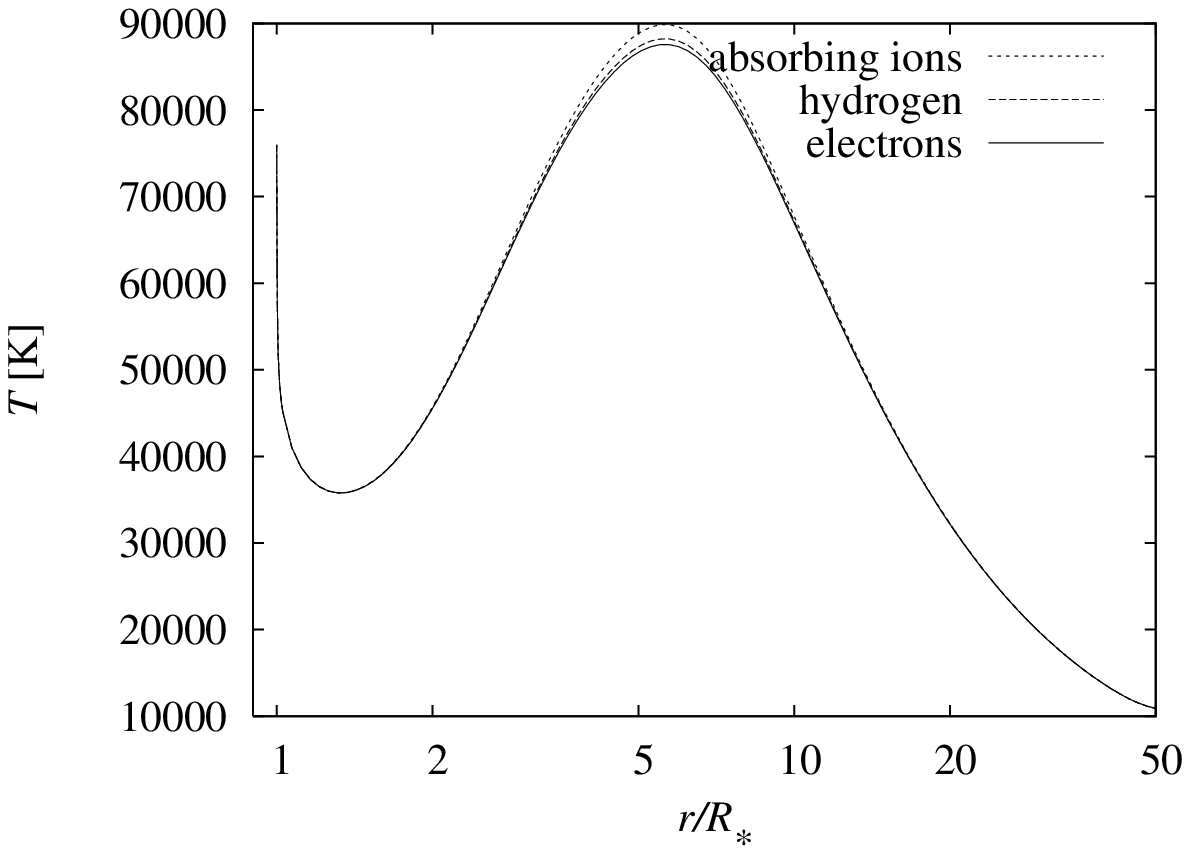}}
\end{center}
\caption{The stellar wind model of star with parameters 
$T_\text{eff}=100\,000\,$K, $M=0.6\,\text{M}_\odot$ and $\log g=6.46$~(CGS).
Velocities of wind components are nearly equal in this case and the
stellar wind is heated in the central parts of the model due to the
frictional heating}
\label{ohrev}
\end{figure}

{\em Frictional heating} is important if the velocity difference is
comparable with averaged thermal speed, $x_{ab}\lesssim1$.
An example of frictionally heated wind model is given in
Fig.\,\ref{ohrev}.

\begin{figure}
\begin{center}
\resizebox{0.49\hsize}{!}{\includegraphics{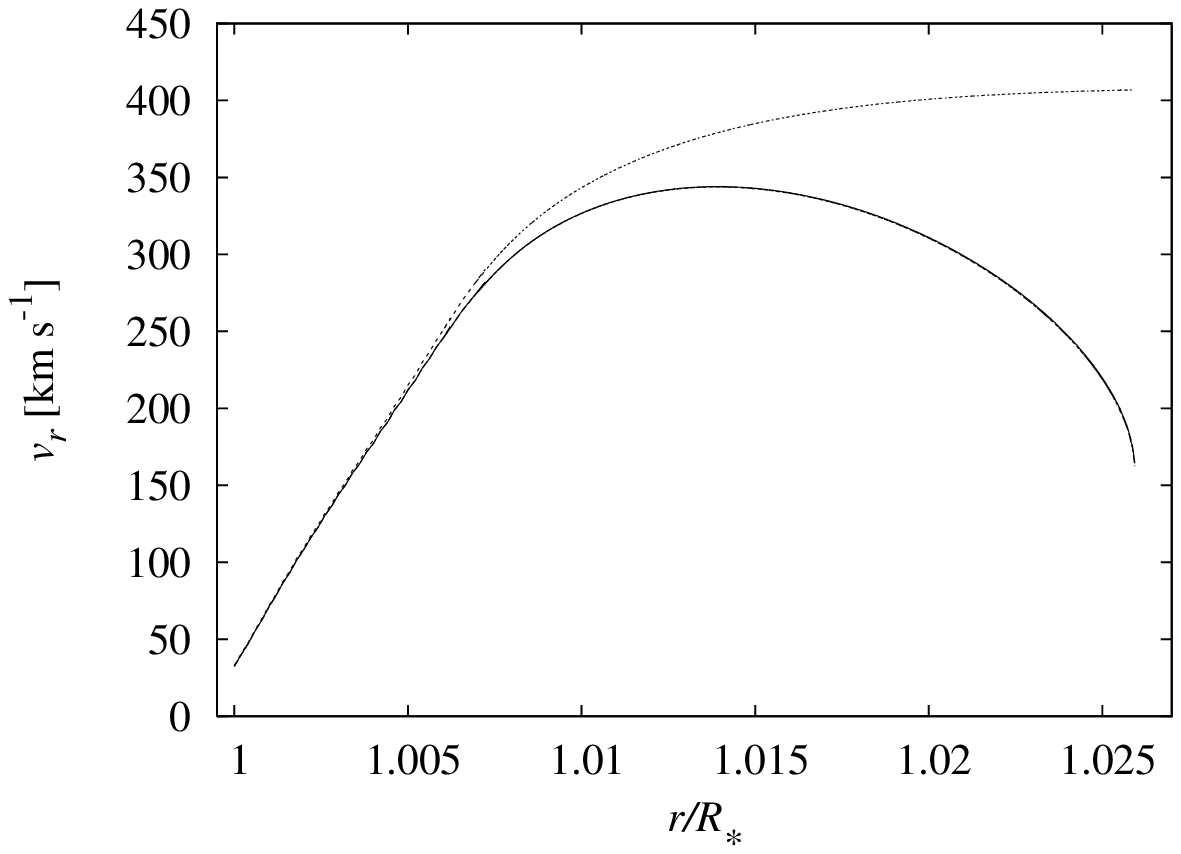}}
\resizebox{0.49\hsize}{!}{\includegraphics{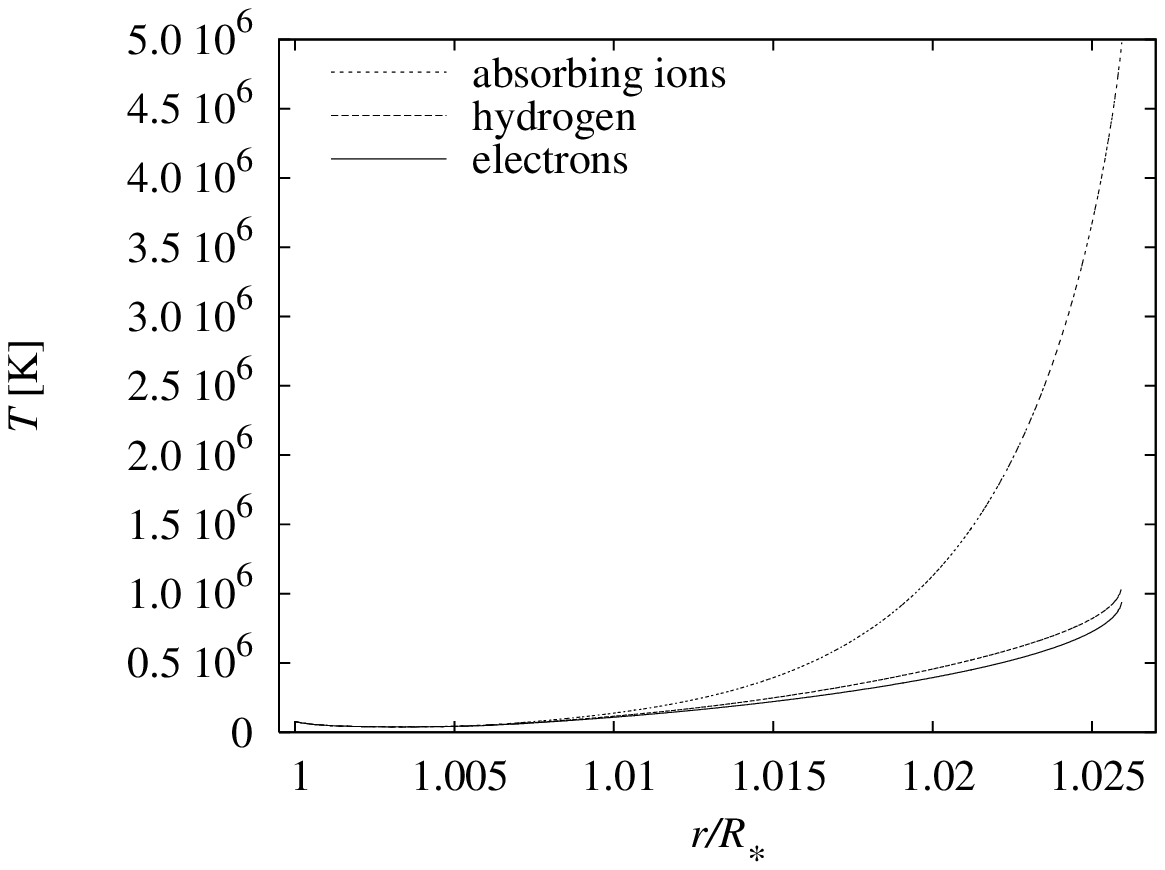}}
\end{center}
\caption{The stellar wind model of a white dwarf with parameters
$T_\text{eff}=100\,000\,$K, $M=0.6\,\text{M}_\odot$ and
$\log g=7.57$~(CGS).
The absorbing component is not able to accelerate hydrogen any more,
hydrogen may fall back onto the stellar surface or may create the
circumstellar shells (Porter \& Skouza~1999).
The stellar wind is significantly frictionally heated in this case.}
\label{hoddel}
\end{figure}

If the relative velocity difference between wind components is higher
than the averaged thermal speed, $x_{ab}\gtrsim1$, {\em the wind
components may decouple}.
An example of model with hydrogen decoupling is given in
Fig.\,\ref{hoddel}.

\section{Regions in $\boldmath{T_\text{eff}/\log g}$ diagram}
%

\begin{figure}
\begin{center}
\resizebox{0.85\hsize}{!}{\includegraphics{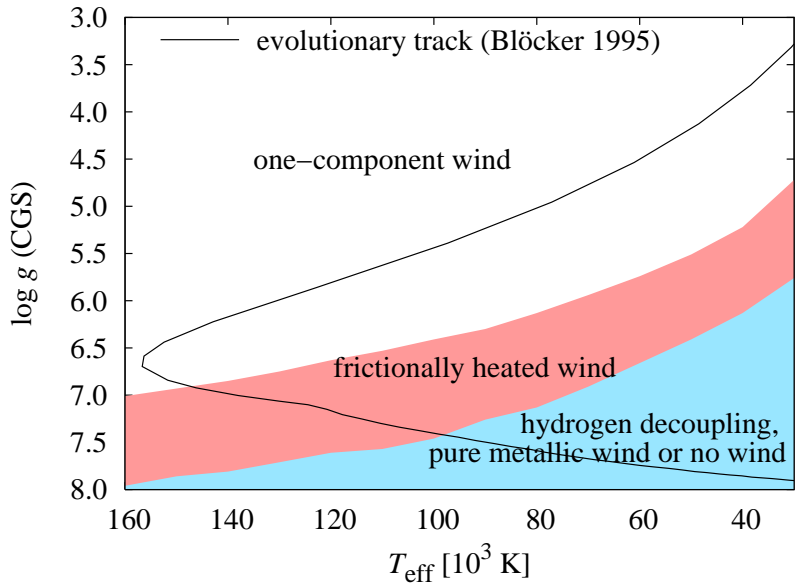}}
\end{center}
\caption{Regions in $T_\text{eff}/\log g$ diagram with different kinds
of stellar wind}
\label{trpaslici}
\end{figure}

Using our wind models, for stars with $M=0.6\,\text{M}_\odot$ we have
derived regions in the $T_\text{eff}/\log g$ diagram with different
types of stellar winds (see Fig.\,\ref{trpaslici}).
Corresponding evolutionary track by Bl\"ocker (1995) of a post-AGB stage
is also plotted in this figure.
Apparently, in the course of the post-AGB evolution also the stellar
wind evolves.
At the initial stages the stellar wind can be regarded as a
one-component.
During the subsequent stellar cooling the stellar wind becomes
frictionally heated and subsequently either hydrogen falls back onto the
stellar surface or pure metallic wind exists.
The coolest stars do not have any wind.

%


\acknowledgements{
We thank the organisers for printing
%
our poster while the original
one was left by one of us (JK) in the train.
This work was supported by grants GA \v{C}R
205/02/0445, 205/03/D020, 205/04/1267.
The Astronomical Institute Ond\v{r}ejov is supported
by projects K2043105 and Z1003909.}

\end{document}